\documentclass[11pt,a4paper]{article}
\usepackage{graphicx}
\usepackage{jcappub}

\def\sun{\hbox{$\odot$}}

\def\la{\mathrel{\mathchoice {\vcenter{\offinterlineskip\halign{\hfil
$\displaystyle##$\hfil\cr<\cr\sim\cr}}}
{\vcenter{\offinterlineskip\halign{\hfil$\textstyle##$\hfil\cr
<\cr\sim\cr}}}
{\vcenter{\offinterlineskip\halign{\hfil$\scriptstyle##$\hfil\cr
<\cr\sim\cr}}}
{\vcenter{\offinterlineskip\halign{\hfil$\scriptscriptstyle##$\hfil\cr
<\cr\sim\cr}}}}}
\def\ga{\mathrel{\mathchoice {\vcenter{\offinterlineskip\halign{\hfil
$\displaystyle##$\hfil\cr>\cr\sim\cr}}}
{\vcenter{\offinterlineskip\halign{\hfil$\textstyle##$\hfil\cr
>\cr\sim\cr}}}
{\vcenter{\offinterlineskip\halign{\hfil$\scriptstyle##$\hfil\cr
>\cr\sim\cr}}}
{\vcenter{\offinterlineskip\halign{\hfil$\scriptscriptstyle##$\hfil\cr
>\cr\sim\cr}}}}}

\title{An upper limit to the secular variation of the gravitational constant 
       from white dwarf stars}
  
\author[a,b]{Enrique Garc\'{\i}a--Berro,}
\author[a,b]{Pablo Lor\'en--Aguilar,}
\author[a,b]{Santiago Torres,}
\author[c,d,1]{Leandro  G.   Althaus\note{Member  of  the  Carrera del
             Investigador  Cient\'{\i}fico  y  Tecnol\'ogico, CONICET,
             Argentina.},}
\author[e,b]{and Jordi Isern}

\affiliation[a]{Departament de F\'\i sica Aplicada, 
                Universitat Polit\`ecnica de Catalunya,\\ 
                c/Esteve Terrades, 5,  
                08860 Castelldefels,  
                Spain}
\affiliation[b]{Institute for Space Studies of Catalonia,\\
                c/Gran Capit\`a 2--4, Edif. Nexus 104,   
                08034  Barcelona, 
	        Spain}
\affiliation[c]{Facultad de Ciencias Astron\'omicas y Geof\'{\i}sicas,  
                Universidad  Nacional de La Plata,\\
                Paseo del  Bosque s/n,  
               (1900) La Plata, 
                Argentina}
\affiliation[d]{Instituto de Astrof\'{\i}sica La Plata, 
                CONICET-UNLP,\\
                Argentina}
\affiliation[e]{Institut de Ci\`encies de l'Espai (CSIC),\\
                Campus UAB, 08193 Bellaterra, 
   	        Spain}

\emailAdd{garcia@fa.upc.edu}

\abstract{A variation of  the gravitational constant over cosmological
          ages modifies  the main  sequence lifetimes and  white dwarf
          cooling ages. Using an state-of-the-art stellar evolutionary
          code we  compute the effects  of a secularly varying  $G$ on
          the main  sequence ages  and, employing white  dwarf cooling
          ages computed  taking into account the effects  of a running
          $G$,  we  place constraints  on  the  rate  of variation  of
          Newton's  constant.   This is  done  using  the white  dwarf
          luminosity  function and  the distance  of the  well studied
          open Galactic  cluster NGC~6791.   We derive an  upper bound
          $\dot  G/G\sim -1.8\times  10^{-12}$~yr$^{-1}$.   This upper
          limit  for  the   secular  variation  of  the  gravitational
          constant compares favorably  with those obtained using other
          stellar evolutionary properties,  and can be easily improved
          if deep  images of the  cluster allow to obtain  an improved
          white dwarf luminosity function.}

\notoc

  
\begin{document}

\maketitle  

   
\section{Introduction}  
\label{intro}  

General Relativity  is currently the preferred  theory of gravitation,
and relies on the  equivalence principle. Consequently, it also relies
on the assumption that the  gravitational constant, $G$, does not vary
with time or  space location.  However, the usual  assumption that $G$
is indeed  constant is  just a hypothesis,  though quite  an important
one,  which  deserves  to   be  explored.   In  fact,  several  modern
grand-unification theories predict  that the gravitational constant is
a slowly varying function of low-mass dynamical scalar fields --- see,
for  instance,  Refs.~\cite{LAea,U,mio}  and references  therein,  for
recent descriptions of the theoretical approaches which can be used to
formally describe  the variation  of fundamental constants.   If these
theories are correct, we expect that the gravitational constant should
experience slow changes over cosmological timescales.

In recent years, several constraints have been placed on the variation
of the fine structure  constant \cite{U,mio}.  This is a controversial
issue,  since  there have  been  recent claims  that  for  a range  of
redshifts ($0.5<z<3.5$) the results are consistent with a time-varying
fine  structure constant \cite{W1,W2,W3},  whereas other  authors have
challenged  these   results  \cite{C04,Sr04,Q04,K05,C05,C06},  or,  at
least,  have  put forth  doubts  on such  a  possible  detection of  a
time-varying $\alpha$  \cite{K10}.  In  sharp contrast with  the vivid
debate about  whether (or  not) there is  evidence for a  varying fine
structure constant, relatively few works  have been devoted to study a
hypothetical variation  of the gravitational  constant.  Probably, one
of the  reasons is the  intrinsic difficulty of measuring  the present
value  of  this  constant  \cite{MTN}.   Actually,  the  gravitational
constant  is the  fundamental  constant  for which  we  have the  less
accurate  determination,  and  the  several  measures  of  $G$  differ
considerably among them.  However,  there are other reasons, being the
weakness   of   the   gravitational   interaction   another   probable
reason. Therefore,  it is  not surprising that  many methods  aimed to
bound  any  hypothetical  variation  of  $G$ have  been  devised.   At
present,  the most  tight constrains  are those  obtained  using Lunar
Laser Ranging and Big  Bang nucleosynthesis.  The most recent analysis
of  the  Lunar  Laser  Ranging  experiments provides  an  upper  bound
$\dot{G}/G = (0.2\pm0.7)\times 10^{-12}$~yr$^{-1}$ \cite{H10}, whereas
the bounds  obtained from Big  Bang nucleosynthesis bounds are  of the
same order of magnitude  $-0.3 \times 10^{-12}$~yr$^{-1} \la \dot{G}/G
\la 0.4\times 10^{-12}$~yr$^{-1}$ \cite{CO4,B05}. However, Lunar Laser
Ranging provides only local limits to the secular rate of variation of
$G$,    whereas    Big     Bang    nucleosynthetic    arguments    are
model-dependent. At intermediate  cosmological ages the Hubble diagram
of  Type Ia  supernovae can  also  be used  to constrain  the rate  of
variation  of  the gravitational  constant,  but  the constraints  are
somewhat  weaker $\dot  G/G\la 1\times  10^{-11}$~yr$^{-1}$  at $z\sim
0.5$~\cite{SNIa,IJMPD}.

White  dwarf  stars allow  to  constrain  in  an independent  way  any
hypothetical variation  of $G$.  The  primary reason for this  is that
the evolutionary timescales of white  dwarfs are very long. Thus, even
small  secular variations  of $G$  become prominent  in the  course of
their lives. However, this is not the only reason whatsoever. In fact,
because  the  inner  cores   of  these  stars  are  almost  completely
degenerate the mechanical structure  of white dwarfs is very sensitive
to the precise value of  $G$, and any, otherwise small, secular change
of  its  value  should  be  quite  apparent  for  long  enough  times.
Morevoer, white dwarfs  do not have nuclear energy  sources, and their
evolution can be well described as a slow cooling process in which the
gravothermal  energy of their  cores is  released through  a partially
degenerate, insulating convective or radiative envelope.  This cooling
process   is  now   very  well   understood  for   sufficiently  large
luminosities,     say     $\log(L/L_{\sun})\ga-4.5$     \cite{nature}.
Additionally, it has been recently shown \cite{gnew} that the specific
rate at which  white dwarfs cool is not only  sensitive to its secular
rate  of  variation,   $\dot  G/G$,  confirming  previous  theoretical
evidence \cite{gold}, but  also to the specific value  of $G$ at which
the white dwarf was born.

In  this paper we  use white  dwarfs to  place an  upper limit  to the
secular rate  of change of the  gravitational constant. To  do this we
use the number counts of white dwarfs of the very old, well-populated,
metal-rich Galactic open cluster NGC~6791. This cluster is so close to
us that  very deep  images can be  obtained \cite{Bedin05}.   This has
allowed to measure with unprecedented accuracy the white dwarf cooling
sequence of  the cluster, and  to obtain reliable number  counts which
ultimately have  allowed to derive  a reliable white  dwarf luminosity
function  \cite{Bedin08a,Bedin08b} ---  that is,  the number  of white
dwarfs as a  function of the magnitude. As it will  be shown below the
shape of  the white dwarf  luminosity function depends on  $\dot G/G$,
and this will allow us to derive an upper bound to the secular rate of
change of $G$.

Our  paper is organized  as follows.   In Sect.   \ref{progenitors} we
describe  the   evolution  of   white  dwarf  progenitors,   while  in
Sect. \ref{wds} we overview the white dwarf cooling tracks employed in
this   work.   We   emphasize   that  these   cooling  sequences   are
state-of-the-art, while  the progenitor evolutionary  sequences with a
varying  $G$ have  been specifically  computed for  this  work.  Sect.
\ref{lf}  is  devoted  to  describe  how the  white  dwarf  luminosity
function of  NGC~6791 was built.  It follows  Sect.  \ref{results}, in
which we  show how $\dot G/G$  can be constrained  using the available
data. Finally, in Sect. \ref{conc} we summarize our major findings, we
discuss its significance and present our concluding remarks.


\section{Evolution of white dwarf progenitors}
\label{progenitors}

As  previously mentioned,  in the  present study  white  dwarf cooling
tracks  that  were derived  following  in  a  self-consistent way  the
evolution of white  dwarfs in the case of a varying  $G$ will be used.
However, a slowly varying $G$ also affects the evolutionary properties
of white dwarf progenitors,  and particularly their ages \cite{Weiss}.
This  is an  important issue  and has  to be  taken into  account when
evaluating the  properties of Galactic  globular or open  clusters. To
assess  the effects  of a  varying $G$  on the  evolutionary  times of
progenitor stars, we have computed  the main sequence evolution of two
model stars  of 1.0  and $2.0\, M_{\sun}$  --- the progenitors  of our
0.525   and  $0.609\,  M_{\sun}$   white  dwarfs,   see  table   1  of
Ref.~\cite{gnew} ---  considering three values for the  rate of change
of $G$, namely $\dot{G}/G=-5 \times 10^{-11}$~yr$^{-1}$, $\dot{G}/G=-1
\times      10^{-11}$~yr$^{-1}$,     and      $\dot{G}/G=-1     \times
10^{-12}$~yr$^{-1}$.   All  the  evolutionary calculations  were  done
using     the    {\tt     LPCODE}     stellar    evolutionary     code
\cite{Althaus10,Renedo10}, appropriately modified to take into account
the effect of a varying $G$.   In particular, we asssume that the rate
of change of  $G$ is so small that the evolution  can be considered as
adiabatic  and  we  simply  allow  $G$  to vary  in  the  equation  of
hydrostatic equilibrium. This approach is fair and has been adopted in
previous  studies  of this  kind  \cite{Weiss}.   Additionally, it  is
important to  realize that the  main sequence evolution  of progenitor
stars  in the  case of  a  varying $G$  is strongly  dependent on  the
initial  value of  $G$  at the  Zero  Age Main  Sequence (ZAMS),  when
hydrogen starts to be burned  in the center of the star.  Accordingly,
for each value of $\dot{G}/G$  we have computed several sequences with
different values of $G_i/G_0$, where $G_i$ stands for the value of $G$
at the ZAMS, and $G_0$ corresponds to the present value of $G$.  It is
important to realize at this  point of our discussion that in previous
works \cite{Weiss}  the important fact  that the initial value  of $G$
should be larger (or smaller,  depending on the adopted value of $\dot
G/G$) than  its present-day value  was not taken into  account.  Thus,
our results  clearly improve the  only existing calculations.   All in
all, we  have computed  19 evolutionary sequences  that are  listed in
Table~\ref{secuencias},  which   cover  the  most   important  stellar
evolutionary phases,  namely the  hydrogen and helium  burning phases,
and the formation  of a carbon-oxygen degenerate core,  but we did not
follow the  thermally pulsing  AGB phase, except  in one case  --- see
below.

\begin{table*}
\caption{Pre-white dwarf evolutionary sequences computed in this work.
         We list  the stellar mass at  the ZAMS (in  solar units) and,
         for each  value of $\dot{G}/G$  (in units of  yr$^{-1}$), the
         initial value of $G$ at  the ZAMS, $G_i/G_0$.  The numbers in
         brackets give  the factor by  which the main sequence  age is
         reduced when a varying $G$ is considered.}
\begin{center}
\begin{tabular}{cccc}
\hline
\hline
\\
$M_{\rm ZAMS}/M_{\sun}$ & \multicolumn{3}{c}{$G_i/G_0$}\\
\cline{2-4}
& 
$\dot{G}/G$=$-5 \times 10^{-11}$ &  
$\dot{G}/G$=$-1 \times 10^{-11}$ &  
$\dot{G}/G$=$-1 \times 10^{-12}$\\
\hline
1.0 &  1.50 ($6.0$)  & 1.15 ($2.3$)  &  1.020 ($1.12$)  \\
1.0 &  1.40 ($4.0$)  & 1.10 ($1.6$)  &  1.015 ($1.07$)  \\
1.0 &  1.30 ($2.6$)  & 1.08 ($1.4$)  &                   \\
\hline
2.0 &  1.50 ($4.8$)  & 1.20 ($1.9$)  &  1.10 ($1.4$)    \\ 
2.0 &  1.40 ($3.4$)  & 1.10 ($1.4$)  &  1.05 ($1.2$)    \\
2.0 &  1.30 ($2.5$)  & 1.05 ($1.2$)  &  1.02 ($1.1$)    \\ 
2.0 &  1.20 ($1.8$)  & 1.03 ($1.1$)  &                   \\
\hline
\hline
\end{tabular}
\end {center}
\label{secuencias}
\end{table*}  

\subsection{Overview of the evolution}

\begin{figure}[t]
\vspace{12cm}    
\includegraphics{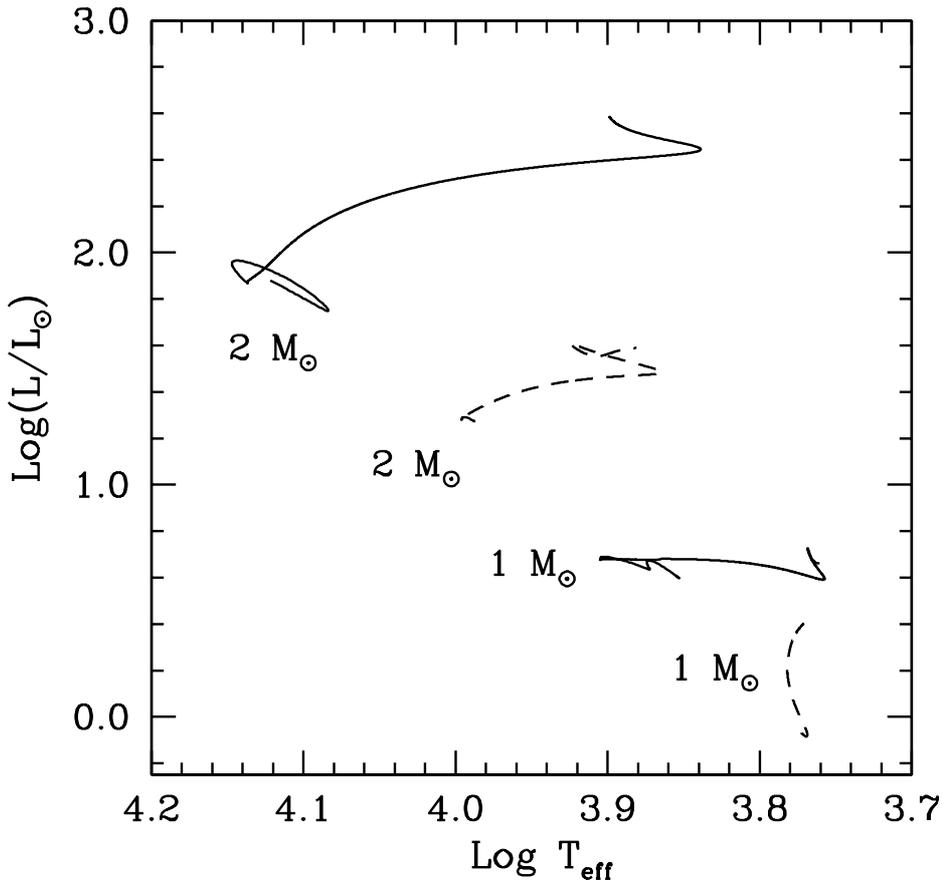}
\caption{Solid lines display  the evolution in the Hertzsprung-Russell
         diagram for the main sequence evolution of stars with 1.0 and
         $2.0\,  M_{\sun}$  assuming  a  rate  of  change  of  $G$  of
         $\dot{G}/G= -5\times10^{-11}$~yr$^{-1}$  and an initial value
         of $G$  at the ZAMS  of $G_i/G_0=1.3$.  Dashed  lines display
         the main sequence evolution for the standard case of constant
         $G$.  See text for details.}
\label{hr}
\end{figure}

Despite the small  rates of change of $G$  adopted here, the evolution
of  white dwarf  progenitor  stars is  severely  modified.  Indeed,  a
varying  $G$ markedly affects  the main  sequence evolution.   This is
illustrated  in Fig.   \ref{hr}, which  shows  the Hertzsprung-Russell
diagram  for   two  stars  with   masses  1.0  and   $2.0\,  M_{\sun}$
respectively, assuming a rate  of change of the gravitational constant
$\dot{G}/G= -5\times10^{-11}$~yr$^{-1}$,  and an initial  value of $G$
at the ZAMS $G_i/G_0=1.3$. In this figure, for the sake of clarity, we
only show the evolution from the onset of core hydrogen burning at the
ZAMS  to  the  stage  at  which the  central  hydrogen  abundance  has
decreased  down   to  $\approx  10^{-9}$  by  mass.    Note  that  the
evolutionary sequences are substantially different from those obtained
in the standard case of constant $G$.  In particular, the evolution in
the case of a varying  $G$ occurs at much higher central temperatures,
which  are a  consequence of  the stronger  gravitational  pull.  This
results in  an enhanced nuclear  burning and, consequently,  in larger
luminosities than  in the  standard case.  This  is in  agreement with
earlier, pioneering  studies \cite{Maeder, Teller},  since from simple
theoretical  grounds it  can  be proven  that  the stellar  luminosity
scales  as a  high  power of  $G$.   Specifically, for  a solar  model
$L\propto  G^7$  \cite{Teller}.   Finally,  we note  that  the  $1.0\,
M_{\sun}$  sequence with  varying $G$  depicted in  Fig.   \ref{hr} is
characterized by an initial  convective core which encompasses $\simeq
30\%$ of  the mass of  the stars.  This  explains the hook  during the
core hydrogen burning phase.

\subsection{Evolutionary timescales}

Of relevance  for the present work  is the impact  on the evolutionary
times.  In particular, the  larger central temperatures of models with
initially  larger values  of $G$  directly translate  in  shorter main
sequence lifetimes, due to the enhanced thermonuclear rates previously
discussed.  The  reduction in the evolutionary  times strongly depends
on the  intial value of $G$ at  the beginning of evolution,  as can be
seen   from  Table~\ref{secuencias},   which,  in   addition   to  the
evolutionary sequences  computed, lists the factors by  which the main
sequence times are  reduced from those predicted by  the standard case
of constant $G$.  Note that  for those sequences with initially larger
values of $G$ the evolutionary timescales are considerably reduced.

The  evolutionary  timescales  can  be modelled  using  rather  simple
arguments.  We  follow closely the treatment  of Ref.~\cite{Weiss}. In
particular, we assume that the  luminosity of a main sequence star can
be factorized as $L\propto f(Y)h(G)$,  where $Y$ is the central helium
abundance. Furthermore,  we assume that  $h(G)\propto G^\gamma$, being
$\gamma$  a constant.   Since the  luminosity is  proportional  to the
helium production it is clear that

\begin{equation}
\frac{dY}{dt}\propto f(Y)h(G),
\end{equation}

\noindent and, hence,  we have that the main  sequence lifetime can be
obtained from

\begin{equation}
\int_{t_0}^{t_0+\tau_{\rm MS}}G(t)^\gamma\, dt\propto\int_{Y_0}^1\frac{dY}{f(Y)},
\end{equation}

\noindent where  $Y_0$ is the  initial helium abundance, and  $t_0$ is
the time at which the star was born. This expression is valid for both
the standard case  and that in which $G$  secularly varies.  Thus, the
left-hand side  of this equation  can be integrated  in the case  of a
constant  $G$, for  which  we  choose the  present-day  value. In  the
following   we  will   assume  that   $\dot  G/G$   remains  constant.
Consequently, if  $\tau_{\rm MS}^0$ is  the main sequence  lifetime in
the standard case,  adopting $t_0=0$, and taking into  account that in
our  evolutionary calculations  we have  adopted a  negative  value of
$\dot G/G$, after some elementary algebra it turns out that:

\begin{equation}
\tau_{\rm MS}=\frac{1}{\gamma\left|\frac{\dot G}{G}\right|}
\ln\left[\gamma\left|\frac{\dot G}{G}\right|\left(\frac{G_0}{G_i}\right)^\gamma
\tau_{\rm MS}^0+1\right].
\label{fit}
\end{equation}

\begin{figure}[t]
\vspace{12cm}    
\includegraphics{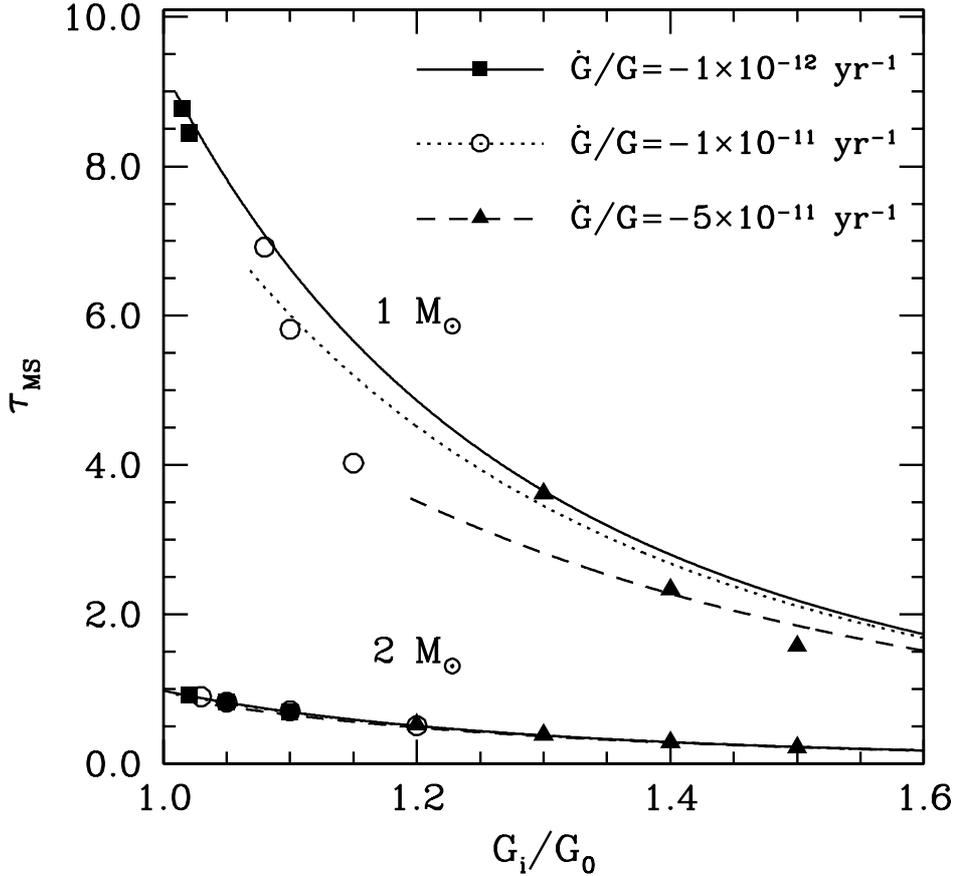}
\caption{Main sequence lifetimes as a function of the initial value of
         the  gravitational constant,  $G_i$, and  different  rates of
         variation  $\dot G/G$.  The  solid, dotted  and dashed  lines
         correspond  to the predictions  of Eq.~(\ref{fit})  for $\dot
         G/G=-1\times    10^{-12}$~yr$^{-1}$,    $\dot    G/G=-1\times
         10^{-11}$~yr$^{-1}$,       and       $\dot       G/G=-5\times
         10^{-11}$~yr$^{-1}$.  Additionally,  the squares, circles and
         triangles correspond to the same set of values of $\dot G/G$,
         and different values of $G_i/G_0$. See text for details.}
\label{Tfit}
\end{figure}

\noindent The  value of $\gamma$ must  be determined from  fits to the
detailed  evolutionary  sequences computed  so  far. Fig.   \ref{Tfit}
shows that this  treatment is relatively accurate, and  --- as it will
be shown in Sect.~\ref{results} ---  enough for our purposes.  In this
figure we show,  for different values of $G_i/G_0$,  the main sequence
lifetimes of two stars of masses $1$ and $2\, M_{\sun}$, respectively,
for different values  of $G_i/G_0$ and $\dot G/G$.   The main sequence
lifetimes in the standard case in  which $G$ is assumed to be constant
are, respectively, 9.46 Gyr for  the $1\, M_{\sun}$ star, and 0.98 Gyr
for the $2\, M_{\sun}$ star.  The solid, dotted and dashed lines show,
respectively,   the   predictions   of  Eq.~(\ref{Tfit})   for   $\dot
G/G=-1\times       10^{-12}$~yr$^{-1}$,       $\dot       G/G=-1\times
10^{-11}$~yr$^{-1}$, and  $\dot G/G=-5\times 10^{-11}$~yr$^{-1}$.  The
results of our evolutionary calculations are shown as circles, squares
and  triangles which  correspond,  respectively, to  the  same set  of
values  of $\dot  G/G$.   The best  fit  corresponds to  $\gamma=3.6$.
Clearly,  the  results  obtained  using Eq.~(\ref{fit})  are  in  good
agreement  with  the evolutionary  results,  especially  for the  more
massive star.  It  should be noted that the  value of $\gamma$ derived
here is  somewhat smaller than that obtained  in previous evolutionary
calculations  \cite{Weiss}. This,  in part,  is due  to  the different
microphysics  used in  the stellar  evolutionary codes.   However, the
largest difference  comes from  the fact that  in our  calculations we
start from  a different value of  $G$, to account  for its present-day
value.

\subsection{Other effects}

Finally,  we have  also explored  the possibility  that a  varying $G$
could change the initial-to-final-mass  relationship.  To this end, we
have  computed the  full evolution  of an  initially  $2.0\, M_{\sun}$
sequence   adopting    $\dot{G}/G=   -5\times10^{-11}$~yr$^{-1}$   and
$G_i/G_0=1.3$. The evolution  was followed all the way  from the ZAMS,
through the hydrogen  and helium core burning phases  to the thermally
pulsing AGB phase, where mass  loss was modeled following the standard
treatment.  We  find that  the mass of  the hydrogen-free core  at the
first thermal pulse, $0.624\, M_{\sun}$, turns out to be substantially
larger  than that obtained  in the  standard case,  $0.52\, M_{\sun}$.
However, the  resulting white dwarf  masses are quite similar  in both
cases. This is  because in the case of the  sequence with varying $G$,
the mass  of the hydrogen-free  core grows only slightly  with further
evolution as  a result of  the larger mass  losses that occur  in this
sequence during  the thermally-pulsing AGB phase.  With  regard to the
core chemical composition of  this sequence, a larger carbon abundance
is expected in  the case of a varying  $G$, resembling the composition
that  results  from  a  $\approx  4.0\,  M_{\sun}$  sequence  computed
assuming  constant $G$, but  this effect  is expected  to be  of minor
importance in the white dwarf cooling phase.

\section{White dwarf cooling tracks}
\label{wds}

In  this study  we  adopt the  most  recent and  reliable white  dwarf
cooling tracks available so  far \cite{gnew}.  These cooling sequences
incorporate the most up-to-date  description of the stellar plasma and
incorporate self-consistently the effects  of a secularly varying $G$.
In particular these cooling sequences consider $^{22}$Ne diffusion and
its  associated   energy  release  \cite{nature,Althaus10,GB08}.   The
energy sources  arising from crystallization  of the white  dwarf core
--- namely,  the release of  latent heat  and of  gravitational energy
associated with carbon-oxygen phase separation \cite{COps,Iea97,Iea00}
--- are  fully taken  into account.   Non-gray model  atmospheres were
used  to provide accurate  outer boundary  conditions for  our models.
Our atmospheres include non-ideal effects in the gas equation of state
and   chemical  equilibrium  based   on  the   occupation  probability
formalism.  They also  consider collision-induced absorption caused by
H$_2$-H$_2$,   H$_2$-He,   and   H-He   pairs,  and   the   Ly$\alpha$
quasi-molecular  opacity that results  from perturbations  of hydrogen
atoms by interactions  with other particles, mainly H  and H$_2$.  Our
white dwarf  cooling tracks also incorporate element  diffusion in the
outer layers, as  well as many other important  physical inputs, which
we do not  detail here for the sake of  conciseness. Instead, we refer
the interested reader to Ref.~\cite{gnew}, were a detailed description
of all them can be found.

The set of  white dwarf cooling tracks employed  here display a marked
dependence of  the white dwarf ages  on the assumed rate  of change of
$G$.  As  a matter of fact,  it turns out that  a secularly decreasing
$G$ accelerates the cooling of the white dwarf. This fact reflects the
energetic demand  required for the  star to expand against  gravity in
response  to  a  smaller  value  of  $G$.   This  dependence  is  more
pronounced   for  massive   white  dwarfs,   owing  to   their  larger
gravitational  forces.   However,  it  is worth  mentioning  that  for
$|\dot{G}/G|\la 1  \times 10^{-12}$~yr$^{-1}$ the  evolution is almost
indistinguishable  from that  of the  standard case  of  constant $G$,
particularly for  low-mass white dwarfs.  Consequently,  this value of
$\dot{G}/G$ constitutes  a lower limit for  the rate of  change of $G$
above which we  can expect that the evolution of  white dwarfs will be
influenced by a varying $G$.

\section{Building the white dwarf luminosity function}
\label{lf}

\begin{figure}[t]
\vspace{14cm}    
\includegraphics{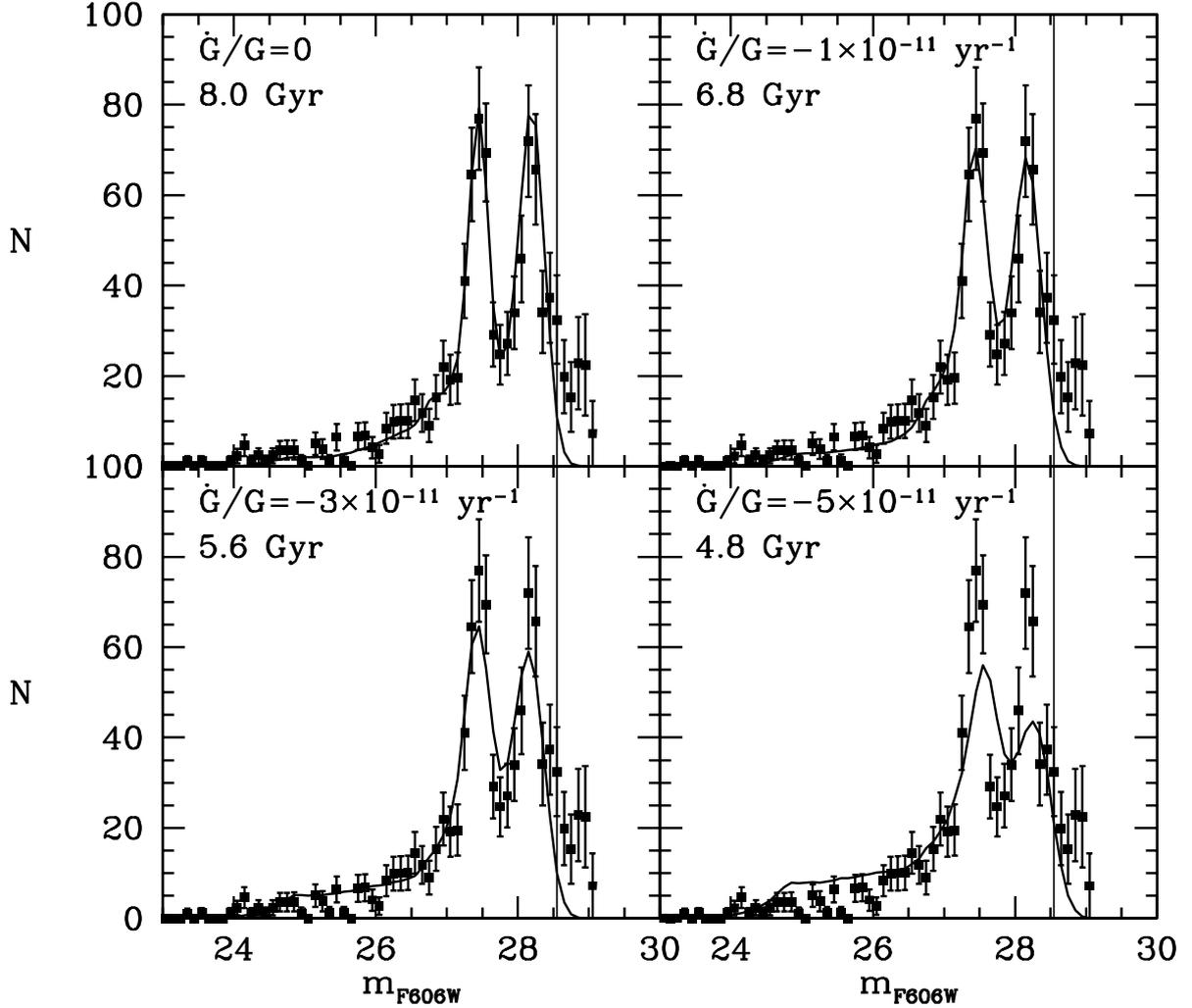}
\caption{The white dwarf luminosity function of NGC~6791 for different
         values   of   $\dot   G/G$.    The  observational   data   of
         Ref.~\cite{Bedin05} are fitted  using the procedures outlined
         in Sect.~\ref{lf}.  The  thin vertical solid line corresponds
         to the magnitude limit beyond which observations are severely
         incomplete.  Note  that the shape of  the luminosity function
         strongly depends on the specific value of $\dot G/G$.}
\label{wdlf}
\end{figure}

The  white dwarf luminosity  function of  NGC~6791 has  been simulated
using a Monte Carlo  technique \cite{MC1,MC2,MC3}.  In our Monte Carlo
simulator synthetic  main sequence stars are  randomly drawn according
to  a Salpeter-like  initial  mass  function that  in  the mass  range
relevant to NGC~6791 white  dwarf progenitors ($M> 1.0\, M_{\sun}$) is
essentially  identical  to  the  ``universal'' initial  mass  function
\cite{Kroupa},  and  a  burst  of  star formation  with  a  small  age
dispersion  of  0.1~Gyr.   At  this  point of  our  discussion  it  is
important  to  mention  that  as   it  will  be  shown  in  detail  in
Sect.~\ref{results}  the age  of the  cluster depends  on  the adopted
value of  $\dot G/G$,  and thus the  time at  which the burst  of star
formation occurred also does.   In accordance with previous studies of
this cluster \cite{Bedin08a, Bedin08b}, we account for a population of
unresolved detached  binary white  dwarfs and we  adopt a  fraction of
binary  systems in  the main  sequence of  54\%.  The  distribution of
secondary masses  is the same adopted in  Ref.~\cite{nature}.  We also
used  an  up-to-date   white  dwarf  initial-final  mass  relationship
\cite{Catalan}.

Given the  age of  the cluster, and  the value  of $\dot G/G$  we know
which was the initial  value of the gravitational constant, $G_i/G_0$,
when the  cluster was  formed. Moreover, once  the time at  which each
synthetic  star  is  randomly   assigned  within  the  burst  of  star
formation, we can compute its associated main sequence lifetime when a
varying  $G$  is adopted  using  Eq.~(\ref{fit})  and up-to-date  main
sequence evolutionary times \cite{WF09}. Thus, we know which synthetic
stars were able  to evolve to the white dwarf stage  and we know which
was the value of $G$  when the corresponding white dwarfs were formed.
Consequently, we  can interpolate their colors  and luminosities using
the theoretical  cooling sequences described in  the previous section.
For unresolved  binary systems we  performed the same  calculation for
the secondary and  we added the fluxes and  computed the corresponding
colors.  The  standard photometric errors in magnitude  and color were
assumed       to      increase      linearly       with      magnitude
\cite{Bedin05,Bedin08a,Bedin08b}.   Finally,  we  added  the  distance
modulus of  NGC~6791, $(m-M)_{\rm F606W}=13.44$, and  its color excess
$E($F606W-F814W$)=0.14$  \cite{Bedin05} to  obtain  a synthetic  white
dwarf  color-magnitude diagram,  and from  it the  corresponding white
dwarf luminosity function was computed.

\section{Results}
\label{results}

\begin{figure}[t]
\vspace{12cm}    
\includegraphics{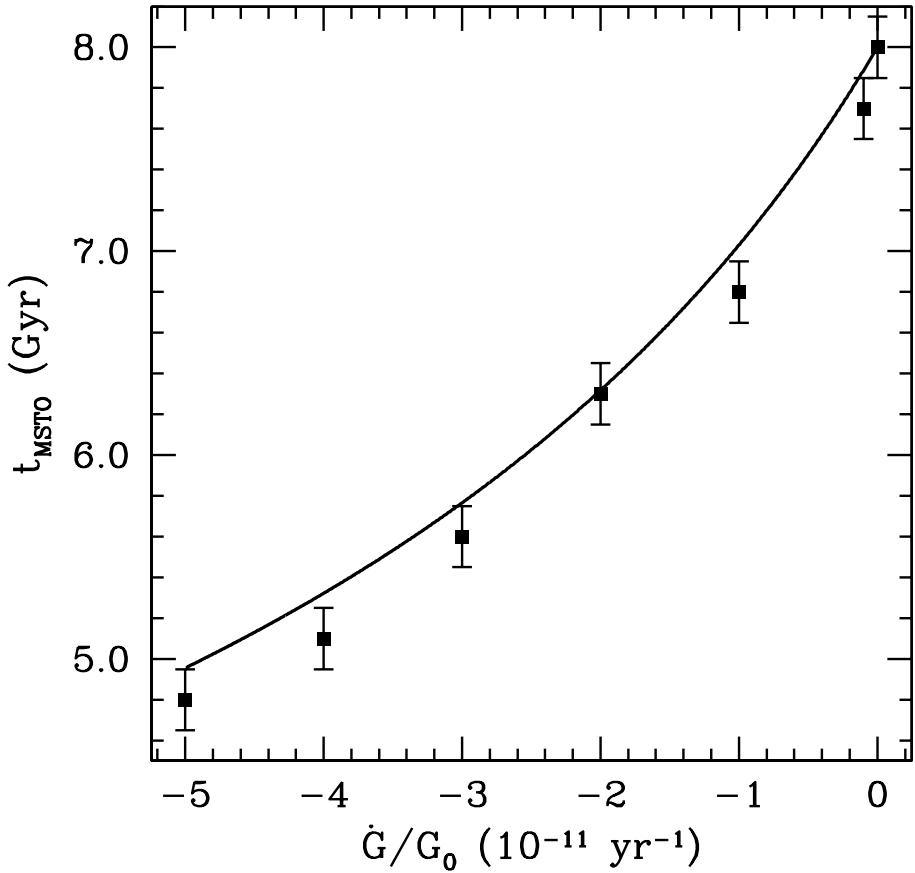}
\caption{Comparison  of  the white  dwarf  cooling  age  and the  main
         sequence turn-off age of NGC~6791 for several values of $\dot
         G/G$.}
\label{TMSTO}
\end{figure}

Using the procedure outlined in  the previous sections the white dwarf
luminosity  function for  an  arbitrary  value of  $\dot  G/G$ can  be
computed.   In Fig.~\ref{wdlf}  we show  several  luminosity functions
corresponding  to increasing absolute  values of  the secular  rate of
change of the gravitational constant.  Each of these panels is clearly
labeled  with  the  specific  value  of $\dot  G/G$  employed  in  the
calculation.  As can  be seen, the white dwarf  luminosity function of
NGC~6791  presents two  peaks.   The bright  peak  corresponds to  the
population of unresolved binary white dwarfs, whilst the faint peak is
due  to finite age  of the  cluster ---  see Ref.~\cite{nature}  for a
detailed explanation of  these features. For the purpose  of this work
it is  important to realize  that white dwarfs with  magnitudes larger
than $\sim 28.7^{\rm mag}$ have not had time to cool enough.  In fact,
the position of  the cut-off --- or, alternatively,  of the faint peak
--- of the white dwarf luminosity function can be used to estimate the
age of the  cluster.  Moreover, this age can be  compared with the age
obtained from the position of the main sequence turn-off.  It has been
recently shown  \cite{nature} that  both ages agree  to a  few percent
when an  appropriate description of  the degenerate plasma  of cooling
white dwarfs  is considered.  Accordingly,  we have fitted the  age of
the cluster  looking for the  age that best  fits the position  of the
faint peak  of the  white dwarf luminosity  function.  These  ages are
also displayed in each of the panels.  Note that as the absolute value
of $\dot G/G$  increases, the age of the cluster  decreases. This is a
consequence of the accelerated cooling of typical white dwarfs --- see
Sect.~\ref{wds}.

Given that  both the main sequence  turn-off ages and  the white dwarf
cooling  ages  depend  on  the  assumed  value of  $\dot  G/G$  it  is
interesting to  compare how the  white dwarf age of  NGC~6791 compares
with the  main sequence  turn-off age, for  different values  of $\dot
G/G$. This can be done inspecting Fig.~\ref{TMSTO}, where we display a
comparison   of  the   age  of   the  main-sequence   turn-off,  which
corrresponds to a mass $M_{\rm MSTO}\simeq 1,137\, M_{\sun}$. The main
sequence lifetime of this star can be scaled using Eq.~(\ref{fit}) and
compared to  the ages  derived from a  fit to the  observational white
dwarf  luminosity  function of  NGC~6791,  which  in  this figure  are
represented using  solid squares.  Note  that the error bars  are very
small, a consequence of the narrowness  of the faint peak of the white
dwarf luminosity function.   As can be seen, the  overall agreement is
excellent, independently of the adopted  value of $\dot G/G$. Thus, we
are  confident  that  the   assumptions  used  to  build  white  dwarf
luminosity functions with a varying $G$ are valid.

\begin{table*}
\caption{Age  of  the  cluster  for different  values  of the rate of 
         change of  $G$, $\dot{G}/G$ (in  units of yr$^{-1}$)  and the
         corresponding probability using the $\chi^2$ test.}
\begin{center}
\begin{tabular}{ccc}
\hline
\hline
\\
$t$ (Gyr) & $\dot{G}/{G}\ {\rm (yr^{-1})}$ & $P$ \\
\\
\hline
8.0 & 0                   & 0.901 \\
7.7 & $-1\times 10^{-12}$ & 0.809 \\
6.8 & $-1\times 10^{-11}$ & 0.795 \\
6.3 & $-2\times 10^{-11}$ & 0.717 \\
5.6 & $-3\times 10^{-11}$ & 0.588 \\
5.1 & $-4\times 10^{-11}$ & 0.264 \\
4.8 & $-5\times 10^{-11}$ & 0.002 \\
\hline
\hline
\end{tabular}
\end {center}
\label{xi2}
\end{table*}  

Now  we turn  our  attention again  to  Fig.~\ref{wdlf}. Clearly,  the
theoretical white  dwarf luminosity  functions strongly depend  on the
adopted  value of  $\dot  G/G$. Actually,  the  agreement between  the
theoretical  luminosity  functions  and  the  observational  data  for
NGC~6791  steadily degrades  as the  absolute value  of $\dot  G/G$ is
increased. This  can be used  to place an  upper limit to the  rate of
variation of the gravitational constant. To quantify the maximum value
of  $\dot G/G$  allowed by  the observational  white  dwarf luminosity
function of  NGC~6791 we have conducted  a $\chi^2$ test,  and we have
found  that  for $\dot  G/G=0$  a  probability  of $P\simeq  0.90$  is
obtained, while for $\dot G/G\simeq-2.5\times 10^{-11}$~yr$^{-1}$ this
value drops to  $P\simeq 0.66$ --- see table~\ref{xi2},  where we show
in addition  to the age of  the cluster obtained from  the position of
the faint peak of the white dwarf luminosity function, the probability
given by the  $\chi^2$ test for several values  of $\dot G/G$.  Hence,
the white dwarf  luminosity function of this open  cluster can be used
to  constrain  the secular  rate  of  variation  of the  gravitational
constant, although  the upper bound  obtained with this method  is not
very  tight.   However, this  upper  limit  compares  well with  other
stellar constraints.   For instance, the upper bound  derived from the
luminosity function  of disk white dwarfs is  $\dot G/G\sim -3.0\times
10^{-11}$~yr$^{-1}$ \cite{gold},  that derived from  globular clusters
is  $\dot  G/G\sim  -3.2\times 10^{-11}$~yr$^{-1}$  \cite{Weiss},  the
pulsating white dwarf  G117-B15A provides an upper limit  of $\dot G/G
\sim -2.5\times 10^{-11}$~yr$^{-1}$ \cite{G117}, while helioseismology
provides a much better (but  purely local) upper bound, $\dot G/G \sim
-1.6\times 10^{-12}$~yr$^{-1}$~\cite{Demarque}.

However, this  is not the best  constraint that can  be obtained using
the  properties of  NGC~6791.   As previously  mentioned,  there is  a
degeneracy between the age of the cluster derived from the termination
of  the cooling  sequence (or,  equivalently, from  the  main sequence
turn-off) and  the upper bound to  $\dot G/G$ obtained  from the white
dwarf luminosity function.  This degeneracy  can be broken if the true
distance of  the cluster  is known by  other means.  Indeed,  if $\dot
G/G\sim -2.5\times  10^{-11}$~yr$^{-1}$ is adopted,  the resulting age
of  the cluster  would be  $\sim 6$~Gyr  --- see  Fig.~\ref{TMSTO} and
Table~\ref{xi2}  ---  but  then  the  position of  the  main  sequence
turn-off  in  the   color-magnitude  diagram  would  be  significantly
different if  the same distance modulus is  adopted.  Accordingly, the
distance modulus necessary to  fit the position in the color-magnitude
diagram  of the  main sequence  turn-off of  the cluster  needs  to be
changed as well.  In particular, if  an age of $\sim 6$~Gyr is adopted
for  NGC~6791, its  distance modulus  needs to  be decreased  by about
0.5$^{\rm mag}$ when  $\dot G/G=0$ is adopted~\cite{nature}.  However,
the distance modulus derived  using a totally independent and reliable
method  (eclipsing binaries) which  does not  make use  of theoretical
models turns  out to be  $13.46\pm 0.1$ \cite{Grundahl}.   Thus, large
errors  in  the  distance   modulus  seem  to  be  quite  implausible.
Accordingly,  such a  small age  for  NGC~6791 (6~Gyr)  can be  safely
discarded.  This argument can be put all the way around, and given the
uncertainty in the distance  modulus ($\simeq 0.1^{\rm mag}$), and the
measured value  of the distance modulus  an upper limit  to $\dot G/G$
can  be  placed.   Specifically,  since  $\Delta  t_{\rm  MSTO}/\Delta
(m-M)_{\rm F606W}\approx  4$~Gyr/mag, the maximum  age difference with
respect to the case in which a constant $G$ is adopted should be $\sim
0.4$~Gyr,  which   translates  into  an  upper   bound  $\dot  G/G\sim
-1.8\times   10^{-12}$~yr$^{-1}$,  see   again   Fig.~\ref{TMSTO}  and
Table~\ref{xi2}.   This upper  limit considerably  improves  the other
existing  upper  bounds  to  the  rate  of variation  of  $G$  and  is
equivalent to the upper limit set by helioseismology.

\section{Conclusions}
\label{conc}

In this paper  we have analyzed the effect  of a varying gravitational
constant on  the evolutionary  properties of white  dwarf progenitors.
We have found that the main sequence lifetimes and other properties of
these stars are  strongly dependent not only on  the rate of variation
of the gravitational constant but also on the adopted initial value of
$G$. Specifically, we  have found that a negative  value of $\dot G/G$
implies  larger temperatures in  the central  regions of  these stars.
This,  in turn,  translates  in enhanced  thermonuclear rates,  larger
luminosities,  and significantly shorter  main sequence  lifetimes, in
agreement with  previous studies \cite{Maeder,Weiss}.   However, these
pioneering studies  only took  into account the  effects of  a varying
gravitational constant  and overlooked the fact that  if $G$ decreases
with time its value should have been larger in the past. Our study has
been the first one to address and quantify this issue. Furthermore, we
have  derived  a simple  analytical  expression  that reproduces  with
reasonable accuracy our numerical results.

With these  tools, and with appropriate white  dwarf cooling sequences
that also  take into  account the  effects of a  varying $G$,  we have
built  up-to-date  white  dwarf  luminosity  functions  for  the  old,
metal-rich,  well-populated, nearby,  Galactic open  cluster NGC~6791.
For  this cluster  we  have a  reliable observational  color-magnitude
diagram  and  a  well  determined  white  dwarf  luminosity  function.
Comparing our results with the  extant observational data we have been
able to derive an upper bound to the secular rate of variation of $G$.
This upper bound, $\dot  G/G\sim -2.5\times 10^{-11}$~yr$^{-1}$, is on
the  order   of  other  constraints  obtained   previously  using  the
evolutionary properties of either  main sequence or white dwarf stars,
and  can  be  easily improved  if  a  reanalysis  of the  white  dwarf
luminosity   function   of  NGC~6791   is   undertaken,  and   smaller
observational   uncertainties  are   achieved  at   magnitudes  within
$26.5^{\rm mag}$ and $28.5^{\rm mag}$ --- something perfectly feasible
using the Hubble  Space Telescope.  Moreover, we also  have found that
combined  observations of the  white dwarf  cooling sequence  and main
sequence stars provide an improved upper bound if the true distance to
the cluster is independently  determined. In particular, using the the
distance  modulus  measured using  eclipsing  binaries  --- a  totally
independent and  reliable method that does not  use theoretical models
--- we  find that  the upper  bound to  the rate  of variation  of the
gravitational  constant can  be improved  to $\dot  G/G\sim -1.8\times
10^{-12}$~yr$^{-1}$, which  compares favorably with the  rest of upper
bounds obtained using stellar evolutionary arguments.

Finally,  we would  like  to mention  that  NGC~6791 is  not the  only
cluster for which we have  white dwarf luminosity functions. There are
other Galactic clusters, either open or globular, for which the Hubble
Space Telescope has  imaged the white dwarf cooling  sequence, and for
which we expect  to have in the near future  very reliable white dwarf
luminosity functions.  These  clusters include M67, NGC~2099, NGC~188,
M4, and  NGC~6397. Their metallicities  are very different of  that of
NGC~6791, and consequently,  appropriate white dwarf cooling sequences
and progenitor  lifetimes should be  computed to obtain  sound results
and  accurate  upper   limits  to  the  rate  of   variation  of  $G$.
Additionally, if reliable upper  bounds are to be obtained independent
distance determinations are needed as well. Nevertheless, our study of
NGC~6791 already paved the way for improved determinations of an upper
limit to the rate of variation of Newton's constant.

\acknowledgments This research was supported by AGAUR, by MCINN grants
AYA2008--04211--C02--01  and  AYA08-1839/ESP,  by  the  ESF  EUROCORES
Program  EuroGENESIS  (MICINN grant  EUI2009-04170),  by the  European
Union   FEDER   funds,  by   AGENCIA:   Programa  de   Modernizaci\'on
Tecnol\'ogica BID 1728/OC-AR, and by PIP 2008-00940 from CONICET.

\bibliographystyle{JHEP}
\bibliography{JCAP}

\providecommand{\href}[2]{#2}\begingroup\raggedright\begin{thebibliography}{10}

\bibitem{LAea}
P.~{Lor{\'e}n-Aguilar}, E.~{Garc{\'{\i}}a-Berro}, J.~{Isern}, and Y.~A.
  {Kubyshin}, {\it {Time variation of G and alpha within models with extra
  dimensions}},  {\em Classical and Quantum Gravity} {\bf 20} (Sept., 2003)
  3885--3896, [\href{http://arxiv.org/abs/arXiv:astro-ph/0309722}{{\tt
  arXiv:astro-ph/0309722}}].

\bibitem{U}
J.~{Uzan}, {\it {The fundamental constants and their variation: observational
  and theoretical status}},  {\em Reviews of Modern Physics} {\bf 75} (Apr.,
  2003) 403--455, [\href{http://arxiv.org/abs/arXiv:hep-ph/0205340}{{\tt
  arXiv:hep-ph/0205340}}].

\bibitem{mio}
E.~{Garc{\'{\i}}a-Berro}, J.~{Isern}, and Y.~A. {Kubyshin}, {\it {Astronomical
  measurements and constraints on the variability of fundamental constants}},
  {\em \aapr} {\bf 14} (Mar., 2007) 113--170.

\bibitem{W1}
J.~K. {Webb}, V.~V. {Flambaum}, C.~W. {Churchill}, M.~J. {Drinkwater}, and
  J.~D. {Barrow}, {\it {Search for Time Variation of the Fine Structure
  Constant}},  {\em Physical Review Letters} {\bf 82} (Feb., 1999) 884--887,
  [\href{http://arxiv.org/abs/arXiv:astro-ph/9803165}{{\tt
  arXiv:astro-ph/9803165}}].

\bibitem{W2}
V.~A. {Dzuba}, V.~V. {Flambaum}, and J.~K. {Webb}, {\it {Space-Time Variation
  of Physical Constants and Relativistic Corrections in Atoms}},  {\em Physical
  Review Letters} {\bf 82} (Feb., 1999) 888--891,
  [\href{http://arxiv.org/abs/arXiv:physics/9802029}{{\tt
  arXiv:physics/9802029}}].

\bibitem{W3}
J.~K. {Webb}, M.~T. {Murphy}, V.~V. {Flambaum}, V.~A. {Dzuba}, J.~D. {Barrow},
  C.~W. {Churchill}, J.~X. {Prochaska}, and A.~M. {Wolfe}, {\it {Further
  Evidence for Cosmological Evolution of the Fine Structure Constant}},  {\em
  Physical Review Letters} {\bf 87} (Aug., 2001) 091301,
  [\href{http://arxiv.org/abs/arXiv:astro-ph/0012539}{{\tt
  arXiv:astro-ph/0012539}}].

\bibitem{C04}
H.~{Chand}, R.~{Srianand}, P.~{Petitjean}, and B.~{Aracil}, {\it {Probing the
  cosmological variation of the fine-structure constant: Results based on
  VLT-UVES sample}},  {\em \aap} {\bf 417} (Apr., 2004) 853--871,
  [\href{http://arxiv.org/abs/arXiv:astro-ph/0401094}{{\tt
  arXiv:astro-ph/0401094}}].

\bibitem{Sr04}
R.~{Srianand}, H.~{Chand}, P.~{Petitjean}, and B.~{Aracil}, {\it {Limits on the
  Time Variation of the Electromagnetic Fine-Structure Constant in the Low
  Energy Limit from Absorption Lines in the Spectra of Distant Quasars}},  {\em
  Physical Review Letters} {\bf 92} (Mar., 2004) 121302,
  [\href{http://arxiv.org/abs/arXiv:astro-ph/0402177}{{\tt
  arXiv:astro-ph/0402177}}].

\bibitem{Q04}
R.~{Quast}, D.~{Reimers}, and S.~A. {Levshakov}, {\it {Probing the variability
  of the fine-structure constant with the VLT/UVES}},  {\em \aap} {\bf 415}
  (Feb., 2004) L7--L11,
  [\href{http://arxiv.org/abs/arXiv:astro-ph/0311280}{{\tt
  arXiv:astro-ph/0311280}}].

\bibitem{K05}
N.~{Kanekar}, C.~L. {Carilli}, G.~I. {Langston}, G.~{Rocha}, F.~{Combes},
  R.~{Subrahmanyan}, J.~T. {Stocke}, K.~M. {Menten}, F.~H. {Briggs}, and
  T.~{Wiklind}, {\it {Constraints on Changes in Fundamental Constants from a
  Cosmologically Distant OH Absorber or Emitter}},  {\em Physical Review
  Letters} {\bf 95} (Dec., 2005) 261301,
  [\href{http://arxiv.org/abs/arXiv:astro-ph/0510760}{{\tt
  arXiv:astro-ph/0510760}}].

\bibitem{C05}
H.~{Chand}, P.~{Petitjean}, R.~{Srianand}, and B.~{Aracil}, {\it {Probing the
  time-variation of the fine-structure constant: Results based on Si IV
  doublets from a UVES sample}},  {\em \aap} {\bf 430} (Jan., 2005) 47--58,
  [\href{http://arxiv.org/abs/arXiv:astro-ph/0408200}{{\tt
  arXiv:astro-ph/0408200}}].

\bibitem{C06}
H.~{Chand}, R.~{Srianand}, P.~{Petitjean}, B.~{Aracil}, R.~{Quast}, and
  D.~{Reimers}, {\it {Variation of the fine-structure constant: very high
  resolution spectrum of QSO HE 0515-4414}},  {\em \aap} {\bf 451} (May, 2006)
  45--56, [\href{http://arxiv.org/abs/arXiv:astro-ph/0601194}{{\tt
  arXiv:astro-ph/0601194}}].

\bibitem{K10}
N.~{Kanekar}, J.~X. {Prochaska}, S.~L. {Ellison}, and J.~N. {Chengalur}, {\it
  {Probing Fundamental Constant Evolution with Neutral Atomic Gas Lines}},
  {\em \apjl} {\bf 712} (Apr., 2010) L148--L152,
  [\href{http://arxiv.org/abs/1003.0444}{{\tt arXiv:1003.0444}}].

\bibitem{MTN}
P.~J. {Mohr}, B.~N. {Taylor}, and D.~B. {Newell}, {\it {CODATA recommended
  values of the fundamental physical constants: 2006}},  {\em Reviews of Modern
  Physics} {\bf 80} (Apr., 2008) 633--730,
  [\href{http://arxiv.org/abs/0801.0028}{{\tt arXiv:0801.0028}}].

\bibitem{H10}
F.~{Hofmann}, J.~{M{\"u}ller}, and L.~{Biskupek}, {\it {Lunar laser ranging
  test of the Nordtvedt parameter and a possible variation in the gravitational
  constant}},  {\em \aap} {\bf 522} (Nov., 2010) L5.

\bibitem{CO4}
C.~J. {Copi}, A.~N. {Davis}, and L.~M. {Krauss}, {\it {New Nucleosynthesis
  Constraint on the Variation of G}},  {\em Physical Review Letters} {\bf 92}
  (Apr., 2004) 171301, [\href{http://arxiv.org/abs/arXiv:astro-ph/0311334}{{\tt
  arXiv:astro-ph/0311334}}].

\bibitem{B05}
C.~{Bambi}, M.~{Giannotti}, and F.~L. {Villante}, {\it {Response of primordial
  abundances to a general modification of $G_{\rm N}$ and/or of the early
  universe expansion rate}},  {\em \prd} {\bf 71} (June, 2005) 123524,
  [\href{http://arxiv.org/abs/arXiv:astro-ph/0503502}{{\tt
  arXiv:astro-ph/0503502}}].

\bibitem{SNIa}
E.~{Gazta{\~n}aga}, E.~{Garc{\'{\i}}a-Berro}, J.~{Isern}, E.~{Bravo}, and
  I.~{Dom{\'{\i}}nguez}, {\it {Bounds on the possible evolution of the
  gravitational constant from cosmological type-Ia supernovae}},  {\em \prd}
  {\bf 65} (Jan., 2002) 023506,
  [\href{http://arxiv.org/abs/arXiv:astro-ph/0109299}{{\tt
  arXiv:astro-ph/0109299}}].

\bibitem{IJMPD}
E.~{Garc{\'{\i}}a-Berro}, Y.~{Kubyshin}, P.~{Lor{\'e}n-Aguilar}, and
  J.~{Isern}, {\it {The Variation of the Gravitational Constant Inferred from
  the Hubble Diagram of Type ia Supernovae}},  {\em International Journal of
  Modern Physics D} {\bf 15} (2006) 1163--1174,
  [\href{http://arxiv.org/abs/arXiv:gr-qc/0512164}{{\tt arXiv:gr-qc/0512164}}].

\bibitem{nature}
E.~{Garc{\'{\i}}a-Berro}, S.~{Torres}, L.~G. {Althaus}, I.~{Renedo},
  P.~{Lor{\'e}n-Aguilar}, A.~H. {C{\'o}rsico}, R.~D. {Rohrmann}, M.~{Salaris},
  and J.~{Isern}, {\it {A white dwarf cooling age of 8 Gyr for NGC 6791 from
  physical separation processes}},  {\em \nat} {\bf 465} (May, 2010) 194--196,
  [\href{http://arxiv.org/abs/1005.2272}{{\tt arXiv:1005.2272}}].

\bibitem{gnew}
L.~G. {Althaus}, A.~H. {C{\'o}rsico}, S.~{Torres}, P.~{Lor{\'e}n-Aguilar},
  J.~{Isern}, and E.~{Garc{\'{\i}}a-Berro}, {\it {The evolution of white dwarfs
  with a varying gravitational constant}},  {\em \aap} {\bf 527} (Mar., 2011)
  A72, [\href{http://arxiv.org/abs/1101.0986}{{\tt arXiv:1101.0986}}].

\bibitem{gold}
E.~{Garc{\'{\i}}a-Berro}, M.~{Hernanz}, J.~{Isern}, and R.~{Mochkovitch}, {\it
  {The rate of change of the gravitational constant and the cooling of white
  dwarfs}},  {\em \mnras} {\bf 277} (Dec., 1995) 801--810.

\bibitem{Bedin05}
L.~R. {Bedin}, M.~{Salaris}, G.~{Piotto}, I.~R. {King}, J.~{Anderson},
  S.~{Cassisi}, and Y.~{Momany}, {\it {The White Dwarf Cooling Sequence in NGC
  6791}},  {\em \apjl} {\bf 624} (May, 2005) L45--L48,
  [\href{http://arxiv.org/abs/arXiv:astro-ph/0503397}{{\tt
  arXiv:astro-ph/0503397}}].

\bibitem{Bedin08a}
L.~R. {Bedin}, I.~R. {King}, J.~{Anderson}, G.~{Piotto}, M.~{Salaris},
  S.~{Cassisi}, and A.~{Serenelli}, {\it {Reaching the End of the White Dwarf
  Cooling Sequence in NGC 6791}},  {\em \apj} {\bf 678} (May, 2008) 1279--1291,
  [\href{http://arxiv.org/abs/0801.1346}{{\tt arXiv:0801.1346}}].

\bibitem{Bedin08b}
L.~R. {Bedin}, M.~{Salaris}, G.~{Piotto}, S.~{Cassisi}, A.~P. {Milone},
  J.~{Anderson}, and I.~R. {King}, {\it {The Puzzling White Dwarf Cooling
  Sequence in NGC 6791: A Simple Solution}},  {\em \apjl} {\bf 679} (May, 2008)
  L29--L32, [\href{http://arxiv.org/abs/0804.1792}{{\tt arXiv:0804.1792}}].

\bibitem{Weiss}
S.~{degl'Innocenti}, G.~{Fiorentini}, G.~G. {Raffelt}, B.~{Ricci}, and
  A.~{Weiss}, {\it {Time-variation of Newton's constant and the age of globular
  clusters.}},  {\em \aap} {\bf 312} (Aug., 1996) 345--352,
  [\href{http://arxiv.org/abs/arXiv:astro-ph/9509090}{{\tt
  arXiv:astro-ph/9509090}}].

\bibitem{Althaus10}
L.~G. {Althaus}, E.~{Garc{\'{\i}}a-Berro}, I.~{Renedo}, J.~{Isern}, A.~H.
  {C{\'o}rsico}, and R.~D. {Rohrmann}, {\it {Evolution of White Dwarf Stars
  with High-metallicity Progenitors: The Role of $^{22}$Ne Diffusion}},  {\em
  \apj} {\bf 719} (Aug., 2010) 612--621,
  [\href{http://arxiv.org/abs/1006.4170}{{\tt arXiv:1006.4170}}].

\bibitem{Renedo10}
I.~{Renedo}, L.~G. {Althaus}, M.~M. {Miller Bertolami}, A.~D. {Romero}, A.~H.
  {C{\'o}rsico}, R.~D. {Rohrmann}, and E.~{Garc{\'{\i}}a-Berro}, {\it {New
  Cooling Sequences for Old White Dwarfs}},  {\em \apj} {\bf 717} (July, 2010)
  183--195, [\href{http://arxiv.org/abs/1005.2170}{{\tt arXiv:1005.2170}}].

\bibitem{Maeder}
A.~{Maeder}, {\it {Four basic solar and stellar tests of cosmologies with
  variable past G and macroscopic masses}},  {\em \aap} {\bf 56} (Apr., 1977)
  359--367.

\bibitem{Teller}
E.~{Teller}, {\it {On the Change of Physical Constants}},  {\em Physical
  Review} {\bf 73} (Apr., 1948) 801--802.

\bibitem{GB08}
E.~{Garc{\'{\i}}a-Berro}, L.~G. {Althaus}, A.~H. {C{\'o}rsico}, and J.~{Isern},
  {\it {Gravitational Settling of $^{22}$Ne and White Dwarf Evolution}},  {\em
  \apj} {\bf 677} (Apr., 2008) 473--482,
  [\href{http://arxiv.org/abs/0712.1212}{{\tt arXiv:0712.1212}}].

\bibitem{COps}
E.~{Garc{\'{\i}}a-Berro}, M.~{Hernanz}, J.~{Isern}, and R.~{Mochkovitch}, {\it
  {Properties of high-density binary mixtures and the age of the universe from
  white dwarf stars}},  {\em \nat} {\bf 333} (June, 1988) 642--644.

\bibitem{Iea97}
J.~{Isern}, R.~{Mochkovitch}, E.~{Garc{\'{\i}}a-Berro}, and M.~{Hernanz}, {\it
  {The Physics of Crystallizing White Dwarfs}},  {\em \apj} {\bf 485} (Aug.,
  1997) 308--312, [\href{http://arxiv.org/abs/arXiv:astro-ph/9703028}{{\tt
  arXiv:astro-ph/9703028}}].

\bibitem{Iea00}
J.~{Isern}, E.~{Garc{\'{\i}}a-Berro}, M.~{Hernanz}, and G.~{Chabrier}, {\it
  {The Energetics of Crystallizing White Dwarfs Revisited Again}},  {\em \apj}
  {\bf 528} (Jan., 2000) 397--400,
  [\href{http://arxiv.org/abs/arXiv:astro-ph/9907077}{{\tt
  arXiv:astro-ph/9907077}}].

\bibitem{MC1}
E.~{Garc{\'{\i}}a-Berro}, S.~{Torres}, J.~{Isern}, and A.~{Burkert}, {\it
  {Monte Carlo simulations of the disc white dwarf population}},  {\em \mnras}
  {\bf 302} (Jan., 1999) 173--188.

\bibitem{MC2}
S.~{Torres}, E.~{Garc{\'{\i}}a-Berro}, A.~{Burkert}, and J.~{Isern}, {\it
  {High-proper-motion white dwarfs and halo dark matter}},  {\em \mnras} {\bf
  336} (Nov., 2002) 971--978,
  [\href{http://arxiv.org/abs/arXiv:astro-ph/0207113}{{\tt
  arXiv:astro-ph/0207113}}].

\bibitem{MC3}
E.~{Garc{\'{\i}}a-Berro}, S.~{Torres}, J.~{Isern}, and A.~{Burkert}, {\it
  {Monte Carlo simulations of the halo white dwarf population}},  {\em \aap}
  {\bf 418} (Apr., 2004) 53--65,
  [\href{http://arxiv.org/abs/arXiv:astro-ph/0401146}{{\tt
  arXiv:astro-ph/0401146}}].

\bibitem{Kroupa}
P.~{Kroupa}, {\it {On the variation of the initial mass function}},  {\em
  \mnras} {\bf 322} (Apr., 2001) 231--246,
  [\href{http://arxiv.org/abs/arXiv:astro-ph/0009005}{{\tt
  arXiv:astro-ph/0009005}}].

\bibitem{Catalan}
S.~{Catal{\'a}n}, J.~{Isern}, E.~{Garc{\'{\i}}a-Berro}, and I.~{Ribas}, {\it
  {The initial-final mass relationship of white dwarfs revisited: effect on the
  luminosity function and mass distribution}},  {\em \mnras} {\bf 387} (July,
  2008) 1693--1706, [\href{http://arxiv.org/abs/0804.3034}{{\tt
  arXiv:0804.3034}}].

\bibitem{WF09}
A.~{Weiss} and J.~W. {Ferguson}, {\it {New asymptotic giant branch models for a
  range of metallicities}},  {\em \aap} {\bf 508} (Dec., 2009) 1343--1358,
  [\href{http://arxiv.org/abs/0903.2155}{{\tt arXiv:0903.2155}}].

\bibitem{G117}
O.~G. {Benvenuto}, E.~{Garc{\'{\i}}a-Berro}, and J.~{Isern}, {\it
  {Asteroseismological bound on $\dot G/G$ from pulsating white dwarfs}},  {\em
  \prd} {\bf 69} (Apr., 2004) 082002.

\bibitem{Demarque}
D.~B. {Guenther}, L.~M. {Krauss}, and P.~{Demarque}, {\it {Testing the
  Constancy of the Gravitational Constant Using Helioseismology}},  {\em \apj}
  {\bf 498} (May, 1998) 871--876.

\bibitem{Grundahl}
F.~{Grundahl}, J.~V. {Clausen}, S.~{Hardis}, and S.~{Frandsen}, {\it {A new
  standard: age and distance for the open cluster NGC 6791 from the eclipsing
  binary member V20}},  {\em \aap} {\bf 492} (Dec., 2008) 171--184,
  [\href{http://arxiv.org/abs/0810.2407}{{\tt arXiv:0810.2407}}].

\end{thebibliography}\endgroup

\end{document}